\documentstyle{article}
\begin{document}
\title{ ON THE DIRECT NUCLEON DECAY OF HIGH-SPIN SUBBARRIER
 SINGLE-PARTICLE STATES IN NEAR-MAGIC NUCLEI}
\author{ G.A.Chekomazov and M.H.Urin}
\date{}
\maketitle
\begin{center}
Moscow Engineering Physics Institute, 115409 Moscow, Russia
\end{center}
\begin{abstract}
The description of the direct nucleon decay of high-spin subbarrier
one-particle states in near-magic nuclei is attempted using a simple optical
model and the simplest version of the coupled-channel approach. The 
branching ratios for the direct decay of the several
single-neutron states in $^{209}Pb$ and $^{91}Zr$ to the ground state and
to the low-lying collective states of
$^{208}Pb$ and $^{90}Zr$, respectively, are evaluated.
Results are compared with 
recent experimental data.
\end{abstract}

 The one-nucleon transfer reactions are the source of a lot of data
on spreading high-lying single-quasiparticle states in medium and heavy
nuclei (see e.g. ref.~\cite{1} and refs. therein). These data have been
successfully described in the case of both near-magic ("hard")~\cite{1,2}
and "soft"~\cite{3,4} spherical nuclei. As it seems, the main interest 
is shifted now to the investigation of decays of the mentioned states.
The experimental data on the direct neutron decays of high-spin
subbarrier single-neutron states in $^{209}Pb$ and $^{91}Zr$
excited by means of the $(\alpha,^3\!\!He)$ reaction
have been reported recently (refs.~\cite{5} and~\cite{6}, respectively).

In the present work we try to describe some of these data:
the relative intensities (branching ratios) for the direct neutron decays of
the  $1k_{17/2}$, $1j_{13/2}$, $2h_{11/2}$ single-neutron states 
in $^{209}Pb$ 
as well as of the  $1i_{13/2}$, $1h_{9/2}$, $1j_{15/2}$
single-neutron states in $^{91}Zr$
to the ground state and to the $3^-$ and $5^-$ low-lying 
excited states of $^{208}Pb$ and $^{90}Zr$.
Respectively, a simple optical model (for description  of 
the decay to the ground state) and the simplest
version of the coupled-channel approach (for description of 
the decay to the
low-lying collective states) have been used for evaluating
the mentioned branching ratios. The results are
compared with the data deduced from the experimental cross sections.

In conformity with recent theoretical considerations of the one-nucleon
transfer reactions (see e.g. ref.~\cite{DWBA})
 we start from the reasonable assumption that there exists
a vertex descriptive of one-nucleon transfer reaction amplitude.
Let $f_{jl}(r,\varepsilon)$ be the  radial part of the vertex
corresponding
to the transferred nucleon having certain values of total $(j)$
and orbital $(l)$ angular momenta. Then the relevant 
part $\sigma_{jl}(\varepsilon)$ of the
energy-averaged inclusive
cross section of the one-nucleon transfer 
to the nucleus in the $0^+$ ground state
is determined by the strength function
corresponding to the vertex $f_{jl}$, which is supposed to be
vanished outside the nucleus:
$$ \sigma_{jl}(\varepsilon)\sim
-\frac{1}{\pi}Im\int f_{jl}^{*}(r,\varepsilon)g_{jl}(r,r';\varepsilon)
f_{jl}(r',\varepsilon)drdr'.\eqno(1)$$
Here, $\varepsilon$ is the energy of the transferred nucleon,
$g_{jl}$ is the radial part of the energy-averaged
single-particle Green function. The simplest description of
this Green function can be obtained within the framework of
an optical model: $g_{jl}\to g^{opt}_{jl}$. 

To illustrate the resonance-like behavior of the cross section (1)
in the vicinity of the energy $\varepsilon_{\lambda}$ of the subbarrier
single-quasistationary state $\lambda\equiv n,j,l$
(~$n\!-\!1$ \ is the radial quantum number)
 we use the pole
representation of the optical-model Green functions 
$g^{opt}_{jl}(r,r';\varepsilon)$, 
which is valid within the nucleus
(see e.g. ref.~\cite{chi0}).
As a result we get:

{\baselineskip=0.4cm
\centerline{$ \sigma_{jl}(\varepsilon)\simeq \sigma_{\lambda}^{(0)}
 S_{\lambda}(\varepsilon), $}
\hbox to 0.99\textwidth{\hfil (2)} 

\centerline{$\displaystyle{ S_{\lambda}(\varepsilon)=-\frac{1}{\pi}Im\int g_{jl}^{opt}(r=r';\varepsilon)dr
\simeq \frac{1}{2\pi}\frac{\Gamma_{\lambda}^{\downarrow}}{
(\varepsilon-\varepsilon_{\lambda})^2+\frac{1}{4}\Gamma_{\lambda}^{\downarrow2}},}$}

}
\medskip
\noindent where $S_{\lambda}$ is the single-particle  strength function~\cite{3};
$\Gamma_{\lambda}^{\downarrow}=2\int W(r,\varepsilon_{\lambda})\vert
\chi_{\lambda}^{(0)}(r)\vert^2dr$ is 
the \ single-particle \ spreading width, $W$ is \ the imaginary
part of \ the optical potential; 
$\sigma_{\lambda}^{(0)}\sim\vert\int f_{jl}(r,\varepsilon)
\chi_{\lambda}^{(0)}(r)dr\vert^2$ is the cross section of one-particle
transfer to the $\lambda$ state, \
 $\chi_{\lambda}^{(0)}$ is the radial part of the wave function 
of the single-particle
quasistationary state. This wave function is normalized 
according to the condition $\int \vert\chi^{(0)}_{\lambda}(r)\vert^2dr=1$.
This integral as the previous ones given in this paragraph is taken up
to the nucleus radius.
 From this point onwards we assume that the
spreading width is much larger than the escape width for the
single-particle direct decay.
	
The energy-averaged reaction amplitude corresponding to the direct nucleon
transfer to a certain single-particle continuum state 
(the target nucleus remains in the $0^+$ ground state) is also determined by 
the vertex mentioned above. Neglecting the fluctuational part of the
energy-averaged cross section, we have:
$$ \sigma_{jl,0^+}^{dir}(\varepsilon)\sim
\big\vert\int f_{jl}(r,\varepsilon)\chi_{\varepsilon jl}^{(+)}(r)
dr\big\vert^2.\eqno(3)$$
Here,  $\chi_{\varepsilon jl}^{(+)}$ is  the radial  part 
of the energy-\-averaged single-\-particle 
wave function of the nucleon-nucleus scattering problem. This wave function is
normalized to the $\delta$-function of energy in the "potential" limit,
in which the coupling
of the single-particle states  to many-particle configurations is neglected.
The optical model can be also used for calculating this wave function.
The branching ratio for the direct nucleon decay to the $0^+$
ground state is defined as the ratio
$$b_{jl,0^+}(\varepsilon)=\frac{\sigma_{jl,0^+}^{dir}(\varepsilon)}{ 
\sigma_{jl}(\varepsilon)} \eqno(4)$$
and can be calculated within the optical model. In the "potential"
limit ($W\to 0$) the branching ratio (4) tends to unity as it
follows from eqs.(1),(3). 

In the vicinity of $\varepsilon_{\lambda}$
the following approximate representation $\chi^{(+)}_{\varepsilon jl}(r)\simeq
A_{\lambda}(\varepsilon)\chi^{(0)}_{\lambda}(r)$ is valid within
the nucleus
(see e.g. ref.~\cite{chi0}). Using the pole representation for
the amplitude $A_{\lambda}(\varepsilon)$ we obtain according to
eqs.(1)-(4):
$$b_{jl,0^+}(\varepsilon)\simeq
\frac{\Gamma_{\lambda}^{\uparrow}(\varepsilon)}{ \Gamma_{\lambda}^{\downarrow}},
\eqno(4')$$
where $\Gamma_{\lambda}^{\uparrow}(\varepsilon)$ 
is the single-particle escape width.
Eq.(4') illustrates the physical meaning of the branching ratio (4)
and shows that this ratio  is expected to be nearly independent of
the vertex $f_{jl}$, provided the vicinity of the relevant single-particle
resonance is considered.
This conclusion is natural, because in view of the absence of the
intermediate structure of the single-particle resonance the
decays of this resonance are independent of the way of
its excitation.

Suppose the target-nucleus excited state can be considered as vibrational with
$L^{\pi},\beta_L$ and $\omega_L$ being the angular
momentum, parity, dynamic deformation parameter and energy of this state,
respectively.
If the particle-phonon coupling is weak (this condition is
fulfilled for near-magic nuclei), the direct part of the
energy-averaged cross section corresponding to the nucleon
transfer into the single-particle continuum state
(the target nucleus remains in the one-phonon state) is described by
the formula, which is the direct generalization of eq.(3).
The generalization is similar to the transition from the
simple optical model to the simplest version of the
coupled-channel approach (see e.g. ref.~\cite{4}) and leads to the
expression:
$$\sigma_{jl,L^{\pi}}^{dir}(\varepsilon)\sim 
\Sigma_{j'l'}\big\vert\int f_{jl}(r,\varepsilon)g_{jl}(r,r';\varepsilon)
v_L(r')\chi_{\varepsilon'j'l'}^{(+)}(r')drdr'\big\vert^2
\frac{\langle jl\Vert Y_L \Vert j'l'\rangle^2 }{ 2j+1},\eqno(5)$$
where
$v_L(r)=(2L+1)^{-1/2}\beta_LR\partial V(r)/\partial r$ is the 
radial part of the one-phonon transition potential,
$R$ is the nuclear radius, $V$ is the single-particle (shell-model)
potential, $\varepsilon'=\varepsilon-\omega_L$,
$\langle jl \Vert Y_L \Vert j'l' \rangle$ is the reduced matrix element.
In the "potential" limit eq.(5) can be directly obtained considering
the particle-phonon interaction as perturbation. By definition,
the ratio of cross sections (5) and (1) is the branching ratio for
the direct nucleon decay to
the low-lying collective states of the target nucleus:
$$b_{jl,L^{\pi}}(\varepsilon)=\frac{\sigma_{jl,L^{\pi}}^{dir}(\varepsilon) }{
\sigma_{jl}(\varepsilon)}
\eqno(6)$$	

In the same approximations used for illustration of branching ratio (4)
by means of eq.(4'), the ratio (6) can be expressed in terms of the
escape widths $\Gamma^{\uparrow}_{\lambda,L^{\pi}}$ for the direct nucleon
decay of the single-particle quasistationary state to the low-lying
collective states:
$$b_{jl,L^{\pi}}\simeq\frac{\Gamma^{\uparrow}_{\lambda,L^{\pi}}}{
\Gamma_{\lambda}^{\downarrow}}\ \ ; \ \
\Gamma^{\uparrow}_{\lambda,L^{\pi}}=2\pi\sum_{j'l'}\big\vert
\int\chi^{(0)}_{\lambda}(r)v_L(r)\chi^{(+)}_{\varepsilon'j'l'}(r)dr\big\vert^2
\frac{\langle jl\Vert Y_L \Vert j'l'\rangle^2 }{ 2j+1}.
\eqno(6')$$	
In accordance with eq.(6') it should be expected that the ratio (6)
is also nearly independent of vertex $f_{jl}$, provided the vicinity
of $\varepsilon_{\lambda}$ is only considered.

In practice branching ratios (4), (6) averaged over finite
excitation energy interval $\Delta$ are considered
(from this point onwards $L^{\pi}$ includes $0^+$):
$$\langle b_{jl,L^{\pi}}\rangle _{\Delta}=\frac{1}{ \Delta}\int
\limits_{(\Delta)}d\varepsilon
b_{jl,L^{\pi}}(\varepsilon).
\eqno(7)$$	
When $\Delta$ contains several single-particle resonances, the averaged
bran\-ching ratio for the direct nucleon decay from the interval $\Delta$
to the $L^{\pi}$ state can be defined as follows:
$$\langle b_{L^{\pi}}\rangle_{\Delta}=\frac{1}{\Delta}{\int\limits_
{(\Delta)}d\varepsilon
\sum_{jl}w_{jl}(\varepsilon)}
b_{jl,L^{\pi}}(\varepsilon),
\eqno(8)$$	
where $w_{jl}(\varepsilon)=\sigma_{jl}(\varepsilon)/\sum_{jl}
\sigma_{jl}(\varepsilon)$ is the
probability of $(jl)$-state excitation. 
These probabilities can be evaluated according to eqs.(2) using
the energy dependences of the cross sections $\sigma_{\lambda}^{(0)}$ and of
the strength functions $S_{\lambda}$ from the DWBA and optical-model calculations,
respectively.
Branching ratios (8) can be
compared with relevant experimental values (deduced after the
elimination of the statistical decay contribution~\cite{5,6}),
provided that probabilities $w_{jl}$ are known.

The parameters  of the optical-model potential as well as 
the parameters of the one-phonon states are the input data for
eveluating the branching ratios considered. They are the same
parameters which are used for the description of the elastic and
inelastic nucleon-nucleus scattering by means of the optical model
and coupled-channel approach.
The choice of the optical-model parameters is performed now within
the dispersive optical-model analysis or the variational
moment approach (see e.g. refs.~\cite{7}).
The analysis of experimental data on both the elastic scattering within
wide energy interval and the single-particle bound states
is needed for the determination of the optical-model parameters.
Nevertheless, we use a simplified way for the choice
of the optical-model potential
as the first step in the analysis of branching ratios
(4), (6)-(8).
Namely, we use "combined" optical-model potential $U_{om}$:
\bigskip

{\baselineskip=0.4cm
\centerline{$ U_{om}(\vec r,\varepsilon)=U(\vec r)+\Delta U(r,\varepsilon) \ \ , \ \
 Re\Delta U(r,\varepsilon)=0.3\varepsilon f_{ws}(r,R,a) \ ,$}
\hbox to 0.99\textwidth{\hfil (9)} 

\centerline{$-Im\Delta U(r,\varepsilon)\equiv W(r,\varepsilon)= - 4a(4.28+0.4
\varepsilon-12.8(N-Z)/A)\partial f_{ws}/\partial r\ \ . $}

}
\noindent
Here, $U(\vec r)=V(r)+V_{so}(\vec r)$ is the  shell-model potential of the  Woods-Saxon type
with parameters taken from ref.~\cite{Chep}:
$$
\begin{array}{cr}
\ \ \ \  V(r)=-53.3(1-0.63(N-Z)/A)f_{ws}\ , \ \ MeV \ \ ; &  \\
\hbox to 0.99\textwidth{\hfil $ V_{so}(\vec r)=14.02(1+2(N-Z)/A)%
(\vec\sigma\vec l \ )\partial f_{ws}/r\partial r\ ,\ \ \ MeV $\ ; \hfil
 (9')}\\
f_{ws}(r,R,a)=[1+exp(r-R)/a]^{-1} \ \ , \ \ R=1.24A^{1/3}\ fm \ \ , \ \
a=0.65 \ fm . \\
\end{array}
$$
The ability of this potential to reproduce the low-energy part of the
experimental single-neutron spectrum for $^{209}Pb$ and
$^{91}Zr$ is presented in Table~1.
The parameters of $\Delta U$ in eq.(9), where the neutron energy $\varepsilon$
is given in $MeV$, are taken from ref.~\cite{8}.
Potential (9),(9') is close to the potential, which has been widely used in
ref.~\cite{4} for the analysis of the low-energy neutron-nucleus
scattering for a great many
spherical nuclei.

Using the potential $U_{om}$ in the limit $W\to0$ we evaluated the
energies $\varepsilon_{\lambda}$ and the escape widths $\Gamma_{\lambda}^{\uparrow}
(\varepsilon_{\lambda})$ for several single-neutron quasistationary states
of $^{209}Pb$ and $^{91}Zr$ (the results are given in Table~2).
These states give the main contribution to the branching ratios
considered.
 The branching ratios
$b_{jl,L^{\pi}}$ and $\langle b_{jl,L^{\pi}}\rangle _{\Delta}$ 
have been evaluated according to (1),(3)--(7) for the 
excitation energy intervals considered
in refs.~\cite{5,6}. 
For these intervals  we have evaluated also the
branching ratios $\langle b_{L^{\pi}}\rangle _{\Delta}$ (8).
The results of the DWBA calculations needed for evaluation of the
probabilities $w_{jl}$ in eq.(8) have been taken from ref.~\cite{9}.
The results of the calculations of the mentioned branching ratios,
the relevant experimental data~\cite{5,6}, the parameters of
low-lying states $L^{\pi}$~\cite{10} are given in Table~3 and
Table~4 for $^{209}Pb$ and $^{91}Zr$, respectively.
The method for calculating the optical-model Green
functions $g^{opt}_{jl}$ is given e.g. in ref.~\cite{4}.
The function $\partial f_{ws}/\partial r$ has been used as $f_{jl}$.
The substitution $\partial f_{ws}/\partial r \to f_{ws}$
changes the $\langle b_{L^{\pi}}\rangle _{\Delta}$ value
less than 20$\%$.
For both nuclei considered
the calculated branching ratios $\langle b_{L^{\pi}}\rangle _{\Delta}$ are in 
qualitative agreement with the experimental values for the
direct neutron decay to the ground and $3^-$ states,
whereas they are in some disagreement for the direct decay to the $5^-$ state.
The possible reason for the disagreement lies in the
description of the $5^-$ states in terms of the dynamic vibration
parameters.

 In the present work the method for evaluating the branching ratios for 
the direct neutron decay of high-spin subbarrier single-particle states in near-magic
("hard") medium and heavy nuclei is given. The method is based on the use of the
optical model and the
simplest version of the coupled-channel approach.
Some applications of the method have been considered.
The next step in development of the method is the use of
modern optical-model potentials and of more advanced versions of the
coupled-channel approach (see e.g. refs.~\cite{7} and~\cite{4},
respectively). 
The use of microscopical calculations for the transition potential
connected with excitation of the low-lying collective states is also
interesting.
The method proposed can be also applied to the description
of the direct proton decay of subbarrier single-proton states.
Relevant experimental data are soon expected~\cite{5}.

The interesting and not fully solved theoretical problem is the
description of the direct nucleon decay of subbarrier single-particle
states in "soft" spherical nuclei where the strong particle-$2^+$-phonon
coupling takes place. The methods for consideration of this coupling~\cite{3,4} 
can be applied to the description of the direct nucleon decay to the ground state and should
be apparently  improved for the description of the direct nucleon decay to the
first $2^+$-state. An attempt of consideration  of this problem
has been undertaken in ref.~\cite{11}. Thus, a comparison of the
calculated branching ratios with experimental ones 
can be a serious test of the theory of the strong particle-$2^+$-phonon
coupling. For this reason, the accumulation of relevant
experimental data seems also to be necessary.

The authors are grateful to S.Fortier and M.N.Harakeh for valuable
discussions and remarks.

The research described in this publication was made possible
in part by Grant No. MQ2000 from the International Science Foundation.
One of the authors (M.H.U.) is grateful to the International Soros
Science Education Program (Professor Subprogram) for support.

\newpage
\begin{center}
{\bf Table 1}
\end{center}
Calculated and experimental~\cite{exen} energies for several single-neutron
bound states of $^{209}Pb$ and $^{91}Zr$.
\begin{center}
\begin{tabular}{|c|c|c|c|c|c|c|c|c|c|}
\hline
 nucleus & \multicolumn{4}{c|}{$^{209}Pb$} & \multicolumn{5}{c|}{$^{91}Zr$} \\
\hline
 $\lambda$  & $2g_{9/2}$  &  $1i_{11/2}$  & $1j_{15/2}$ & 
$2g_{7/2}$ & $ 2d_{5/2}$ & $3s_{1/2}$& $2d_{3/2}$ & $ 1h_{11/2}$ & $1g_{7/2}$ \\
\hline
 $\varepsilon^{calc}_{\lambda},\ (-1) MeV$ & 4.27 & 2.53 & 3.03 & 0.18 & 
6.79 & 4.96 & 3.89 & 2.73 & 4.77 \\
\hline
 $\varepsilon^{exp}_{\lambda},\ (-1) MeV$ & 3.94 & 3.16 & 2.52 & 1.45 &
7.19 & 5.99 & 5.16 & 5.03 & 5.00 \\
\hline
\end{tabular}
\end{center}

\begin{center}
{\bf Table 2}
\end{center}
Calculated energies and escape widths for several single-neutron
quasistationary states of $^{209}Pb$ and $^{91}Zr$.
\begin{center}
\begin{tabular}{|c|c|c|c|c|c|c|}
\hline
 nucleus & \multicolumn{3}{c|}{$^{209}Pb$} & \multicolumn{3}{c|}{$^{91}Zr$} \\
\hline
 $\lambda$  & $2h_{11/2}$  &  $1k_{17/2}$  & $1j_{13/2}$ & 
$1i_{13/2}$ & $1h_{9/2}$ & $1j_{15/2}$ \\
\hline
 $\varepsilon_{\lambda},\ MeV$  &   2.680   &  4.989  &   8.056   & 
 7.319 & 7.349 & 17.320 \\
\hline
  $\Gamma^{\uparrow}_{\lambda}(\varepsilon_{\lambda}),\ keV$  &   105 
  &  2.1  &   160   & 185 & 660 & 3180 \\
\hline
\end{tabular}
\end{center}


\begin{center}
{\bf Table 3}
\end{center}
Calculated and experimental branching ratios for the direct neutron
decay of $^{209}Pb$ from the excitation energy intervals $\Delta_1$=8.5--10\ MeV 
and $\Delta_2$=10--12\ MeV (upper and lower lines, respectively).
\begin{center}
\begin{tabular}{|c|c|c|c|c|c|c|c|}
\hline
$L^{\pi}$ &  $\omega_L$ &  $\beta_L$  & \multicolumn{3}{c|}%
{$\langle b_{jl,L^{\pi}}\rangle_{\Delta},\ \%$} & \multicolumn{2}{c|}%
{$\langle b_{L^{\pi}}\rangle_{\Delta}, \%$}  \\
\cline{4-8}
  & $MeV$ &  & $2h_{11/2}$  & $1k_{17/2}$  & $1j_{13/2}$  & calc  & exp  \\
\hline
  &  &  &    27        &   0.07       &   0.57       & 0.41  & 0.49  \\
$0^+$ & 0  &  0  & & & & &  \\
 &  &  &    35        &   0.35       &   1.9        & 0.81  & 0.36  \\
\hline
  &  &  &    1.10      &    0.98      &    0.69      & 0.92  & 2.1   \\
$3^-$ &  2.6   &  0.1   & & & & & \\
  & &   &    0.81      &    1.30      &    0.76      & 1.13  & 1.5    \\
\hline
  &  &  &    0.46      &    0.11      &    0.38      & 0.16  & 7.0    \\
$5^-$ &   3.2   &   0.05   & & & & & \\
  &  &  &    0.30      &    0.25      &    0.31      & 0.27  & 1.3    \\
\hline
\end{tabular}
\end{center}

\begin{center}
{\bf Table 4}
\end{center}
Calculated and experimental branching ratios for the direct neutron
decay of $^{91}Zr$ from the excitation energy interval
$\Delta$=11--15 MeV.
\begin{center}
\begin{tabular}{|c|c|c|c|c|c|c|c|}
\hline
$L^{\pi}$ &  $\omega_L$ &  $\beta_L$  & \multicolumn{3}{c|}%
{$\langle b_{jl,L^{\pi}}\rangle_{\Delta},\ \%$} & \multicolumn{2}{c|}%
{$\langle b_{L^{\pi}}\rangle_{\Delta}, \%$}  \\
\cline{4-8}
  & $MeV$ &  & $1i_{13/2}$  & $1h_{9/2}$  & $1j_{15/2}$  & calc  & exp  \\
\hline
$0^+$ & 0 & 0 &    1.1        &   4.5       &   0.2       & 1.6  & 1.4  \\
\hline
$5^-$ &  2.31  &  0.08 &  0.5      &    0.2      &    0.1      & 0.4  & 0.   \\
\hline
$3^-$ &  2.75  &  0.2 &  3.1      &    0.9      &    0.1      & 2.1  & 2.4   \\
\hline
\end{tabular}
\end{center}

\begin{thebibliography}{99}
\bibitem{1}  S.Gales, Ch.Stoyanov and A.I.Vdovin,  Phys.Rep. 166 (1988) 125
\bibitem{2}  A.I.Vdovin, V.V.Voronov, V.G.Soloviev and
Ch.Stoyanov, \\ Sov.J.Part.Nucl. 16 (1985) 105
\bibitem{3}  S.E.Muraviev, B.A.Tulupov and M.H.Urin, Z.Phys. A341 (1992) 383
\bibitem{4}  V.V.Samoilov and M.H.Urin, Nucl.Phys. A567 (1994) 237 
\bibitem{5}  S.Gales, Nucl.Phys. A569 (1994) 393c; 
\newline     D.Beaumel et al., Phys.Rev. C49 (1994) 2444 
\bibitem{6}  S.Fortier, Decay modes of high-lying single-particle
states, in: Proc. IV International Conference on Selected Topics in
Nuclear Structure (Dubna, Russia, July 5-9, 1994), to be published
\bibitem{DWBA} T.Udagawa, Y.J.Lee and T.Tamura, Phys.Rev. C39 (1989) 47
\bibitem{chi0} M.H.Urin, Sov.J.Part.Nucl. 8 (1977) 331
\bibitem{7}  C.H.Johnson, D.J.Horen and C.Mahaux, Phys.Rev. C36 (1987) 2252;
\newline     C.Mahaux and R.Sartor, Nucl.Phys. A493 (1989) 157 
\bibitem{Chep} V.A.Chepurnov, Sov.J.Nucl.Phys. 6 (1967) 696
\bibitem{8}  J.Rapaport, V.Kulkarni and R.W.Finlay, Nucl.Phys. A330 (1979) 15
\bibitem{9}  P.D.Kunz, code Dwuck4, unpublished
\bibitem{10} H.P.Morsch et al., Phys.Rev. C22 (1980) 489;
\newline     W.S.Gray, R.A.Kenefick, J.J.Kraushaar and G.R.Satchler,
Phys.Rev. 142 (1966) 735
\bibitem{11} S.E.Muraviev and M.H.Urin, Phys.Lett. B262 (1991) 185
\bibitem{exen} S.Fortier et al., Phys.Rev. C41 (1990) 2689
\end{thebibliography}
\end{document}